# A novel brain registration model combining structural and functional MRI information

Baolong Li[1], Yuhu Shi[1]*, Lei Wang[1], Weiming Zeng[1], and Changming Zhu[1]

[1] Information Engineering College, Shanghai Maritime University, 1550 Harbor Avenue, Pudong, Shanghai, 201306, China

* **Corresponding author**: College of Information Engineering, Shanghai Maritime University, Shanghai, 201306, China; E-mail address: syhustb2011@163.com.

**Abstract**: Although developed functional magnetic resonance imaging (fMRI) registration algorithms based on deep learning have achieved a certain degree of alignment of functional area, they underutilized fine structural information. In this paper, we propose a semi-supervised convolutional neural network (CNN) registration model that integrates both structural and functional MRI information. The model first learns to generate deformation fields by inputting structural MRI (T1w-MRI) into the CNN to capture fine structural information. Then, we construct a local functional connectivity pattern to describe the local fMRI information, and use the Bhattacharyya coefficient to measure the similarity between two fMRI images, which is used as a loss function to facilitate the alignment of functional areas. In the inter-subject registration experiment, our model achieved an average number of voxels exceeding the threshold of 4.24 is 2248 in the group-level t-test maps for the four functional brain networks (default mode network, visual network, central executive network, and sensorimotor network). Additionally, the atlas-based registration experiment results show that the average number of voxels exceeding this threshold is 3620. The results are the largest among all methods. Our model achieves an excellent registration performance in fMRI and improves the consistency of functional regions. The proposed model has the potential to optimize fMRI image processing and analysis, facilitating the development of fMRI applications.

# 1 Introduction

Magnetic resonance imaging (MRI) technology is an important medical imaging technique for exploring the human brain. This technology has been favored by many scholars in brain research because of the advantage of being non-invasive. Depending on different MRI techniques, different MRI images of the brain can be obtained, such as T1-weight MRI (T1w-MRI), T2-weight MRI (T2w-MRI), functional MRI (fMRI), etc. Among them, T1w-MRI and T2w-MRI have high resolution and rich texture information [1], which can clearly show the fine structural information of the brain. In contrast, fMRI measures brain activity by detecting correlated changes in blood flow, resulting in a blood oxygen level-dependent signal [2]. It has a lower resolution and mainly carries a time series of functional signals.

The raw images obtained by the machine must undergo a series of preprocessing to facilitate further analysis. The image registration is a fundamental task in MRI preprocessing. It aims to find the transformation parameters (or deformation fields) between a pair of images, so that one (moving or source) image is warped and aligned to another (fixed or reference) image. Specifically, the registration is usually used for spatial normalization in image preprocessing, which aligns the MRI scans of subjects to a common space, such as Talairach space [3] or Montreal Neurological Institute (MNI) space [4]. After registration, all subjects can be in the same space for further group analysis and follow-up studies.

Due to the low spatial resolution of fMRI, the registration of fMRI data is usually performed in a stepwise manner by their co-registration structural MRI (sMRI) [5-7]. For example, in the spatial normalization of the preprocessing tool DPARSF [6], the brain fMRI is affine registered with the corresponding T1w-MRI. Then the MNI spatial template is selected as the reference image, and the T1w-MRI of the subject is spatially transformed using the DARTEL [8] method. The same image transformation parameters are later applied to the fMRI corresponding to them to complete the spatial normalization. In addition, the same registration steps can be completed in the Ants Toolkit [7]. Most of the traditional registration algorithm studies have focused on T1w-MRI, which are based on linear spatial transformation [9] and nonlinear spatial transformation methods [8, 10-13]. They achieved the desirable results, but each registration requires a large number of iterations to generate a specific deformation field for a pair of images, which is very time-consuming when processing a large number of images.

With the development of deep learning, researchers have gradually turned to use convolutional neural networks (CNN) for registration studies. In supervised methods, deformation fields are often used as labels for training. Ground truth deformation fields are either synthetically generated or generated by traditional methods. For example, Yang et al. [14] proposed a supervised convolutional neural network for predicting the deformation fields generated by the LDDMM algorithm from image blocks. Sokotti et al. [15] trained the registration convolutional network using synthetic displacement fields. Due to the limitations of supervised methods, the focus of research has shifted to unsupervised or semi-supervised methods that require no ground truth deformation fields [16-22]. For example, De Vos et al. [16] proposed a deep learning framework for unsupervised affine and deformable image registration. Balakrishnan et al. [17] proposed a learning framework for deformable medical image registration based on U-Net and introduced anatomically segmented images to aid in registration.

Recent studies have shown that functionally defined regions are not consistently located relative to anatomical landmarks [23], the alignment of structural regions does not imply the establishment of functional consistency. Therefore, some direct registration methods for fMRI have been proposed [1, 24-28], which extract valid functional features from fMRI grayscale images and perform alignment based on these functional features. For example, Conroy et al. [25] proposed that functional features extracted from whole-brain functional connectivity (FC) could be used to assist in registration. Later, Jiang et al. [26] proposed to used local FCs with better anti-perturbation performance to accomplish the registration. With the emergence of registration algorithms for deep learning, Zhu et al. [1] proposed an unsupervised fMRI registration algorithm based on U-Net, and also proposed a semi-supervised algorithm based on it using gray matter and white matter as auxiliary information. This was followed by Zhu et al. [27], who proposed a 30-layer constant mapping cascade network for fMRI registration. Although the above algorithms utilized FC to achieve alignment of functional regions to some extent, they did not take advantage of the fine structural information of T1w-MRI, nor did they discuss the alignment performance of structural regions in depth.

Based on the above problems, we proposed a semi-supervised CNN registration model by combining structural and functional MRI information. The model used the fine structural information of T1w-MRI to generate the deformation field, and utilized the fMRI local FC as auxiliary information to achieve the structural spatial correspondence, which enhancing the statistical significance of the functional regions. Then, the network parameters are learned and continuously optimized though training sets, in which the GPU is used for accelerating computation and achieving high-quality alignment of fMRI images.

## 2 Materials and methods

In this section, we will provide a detailed description of how to fuse the information of multimodal MRI images and register moving subjects to fixed subjects. The contents of CNN and spatial transformer network are first introduced, and then the methods proposed in this paper are presented.

## 2.1 Convolution neural network

We use the three-dimensional (3D) U-Net model [29, 30] in our CNN model. 3D U-Net is a standard fully convolution end-to-end neural network that generates a set of feature maps with different resolutions. The shallow feature map has high resolution and rich detail information such as texture information. While the deeper feature map has low resolution, the receptive field is getting larger and its semantic information is richer. The excellent ability of 3D U-Net to capture local and global contextual information is very suitable for the field of medical image.

Moreover, because T1w-MRI has clear structural details, we use moving T1w-MRI ($M_{T1}$) and fixed T1w-MRI ($F_{T1}$) as the input of 3D U-Net. The network captures the structural and semantic information of the brain, and generates a T1w-MRI deformation field ($\phi_{T1}$) according to the voxel position correspondence between the two images.

## 2.2 Spatial transformation network

After the 3D U-Net operation in the previous section, we obtain the $\phi_{T1}$. And the fMRI deformation field ($\phi_f$) is also obtained by down sampling. Next, the two deformation fields are respectively applied to the corresponding image to be warped nonlinearly through the spatial transformer network (STN) [31].

For the T1w-MRI, the $\phi_{T1}$ is applied to the $M_{T1}$, symbolically denoted as $M_{T1} \circ \phi_{T1}$, and the warped T1w-MRI ($R_{T1}$) is obtained after registration. The spatial transformation function is expressed as follows, and the intensity of each voxel position p of the output image is calculated by linear interpolation:

$$R_{T1} = M_{T1} \circ \phi_{T1}(p)$$
$$= \sum_{q \in Z(\hat{p})} M_{T1}(q) \prod_{d \in \{x,y,z\}} (1 - |\hat{p}_d - q_d|) \qquad (1)$$

where $\hat{p} = p - u_{T1}(p)$ represents the intensity of position $p$ after the transformation of the displacement field $u_{T1}$. $q \in Z(\hat{p})$ denotes $q$ as the voxel position of the region adjacent to $\hat{p}$. And $d$ represents the three directions of x, y, and z in 3D space.

Similarly, the $\phi_f$ is then applied to the corresponding fMRI moving image ($M_f$) for the pair of subjects selected above. The corresponding warped fMRI images ($R_f$) are obtained by linear interpolation.

$$R_f = M_f \circ \phi_f(n)$$
$$= \sum_{m \in Z(\hat{n})} M_f(m) \prod_{d \in \{x,y,z\}} (1 - |\hat{n}_d - m_d|) \qquad (2)$$

where $\hat{n} = n - u_f(n)$ represents the position $\hat{n}$ after displacement field $u_f$ transformation, and $m \in Z(\hat{n})$ means $m$ is the voxel position of adjacent regions of $\hat{n}$. $d$ represents the three directions of x, y, and z in 3D space. The spatial transformation function can calculate the gradient or sub-gradient, so it can back-propagate the error during optimization

## 2.3 Semi-supervised registration model combining structural and functional MRI information

We proposed a semi-supervised model based on CNN that fuses structural and functional information based on the network structures used in the previous two sections, as shown in **Fig. 1**. It takes T1w-MRI image pair ($M_{T1}$ and $F_{T1}$) and fMRI image pair ($M_f$ and $F_f$) as input and outputs the warped images ($R_{T1}$ and $R_f$). The training process of the whole model is outlined as follows:

Step 1: $M_{T1}$ and $F_{T1}$ are used as the input part of U-Net, and $\phi_{T1}$ and $\phi_f$ are generated through the network.

Step 2: $M_{T1}$ and $M_f$ are warped by the spatial transformer network to obtain the $R_{T1}$ and $R_f$.

Step 3: Use the similarity metric to calculate the loss function between the result and our target images, and then combining these loss functions constitutes a complete semi-supervised loss to assist registration.

Step 4: The parameters of the network are updated by the optimizer gradient. The model is iteratively trained on the training set and continues to update the network parameters minimizing the loss to obtain the final model parameters.

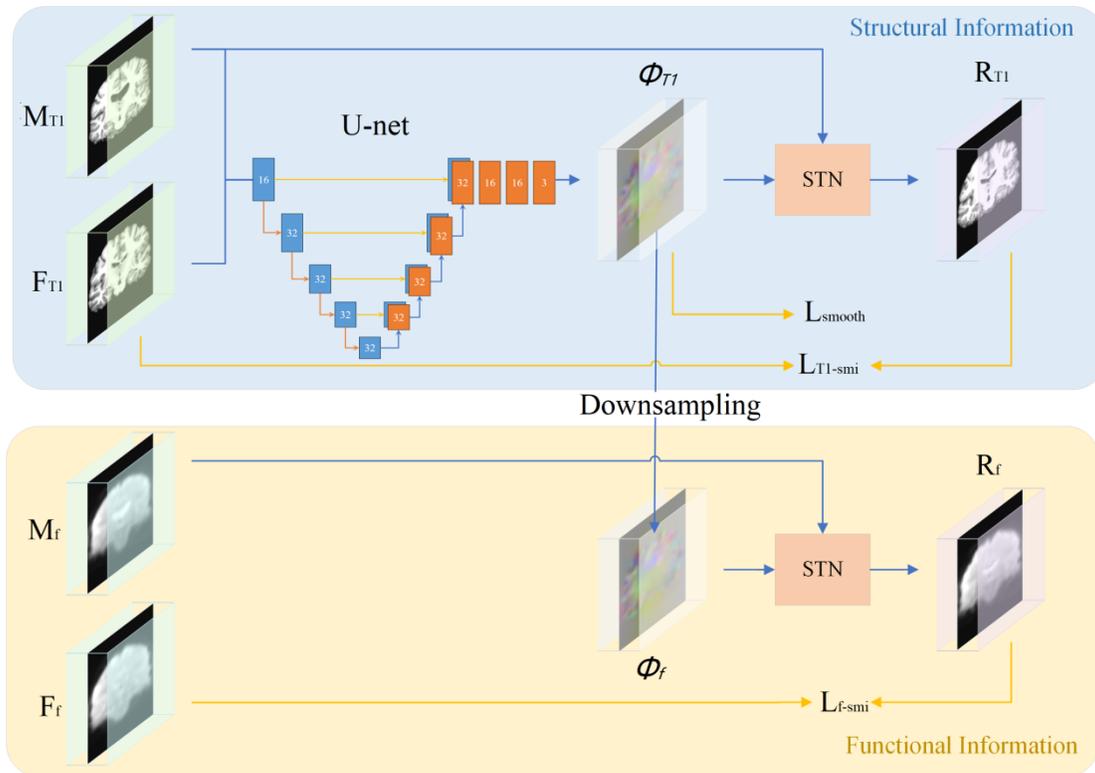

**Fig. 1** The framework of semi-supervised registration model combining structural and functional information of magnetic resonance imaging.

**Fig. 2** depicts the structure of 3D U-Net, symbolically denoted as $g_\theta$, including the encoder and decoder. Two 3D images connected as 2 channels $M_{T1}$ and $F_{T1}$, which are accepted as inputs in the encoder. Down-sampling is achieved through a convolutional kernel with a stride of 2 and a size of 3, which allows the spatial dimension of each layer to be reduced by half. Each convolution operation is followed by the

LeakyReLU activation function layer with parameter 0.2. The convolution layer captures the hierarchical features of the input structure image pair for estimating the fine deformation field $\phi_{T1}$. In order to comprehensively consider the registration accuracy and performance, four-step convolutions are performed in our 3D U-Net structure, and the smallest layer is convolved on 1/16 of the volume of the input image size, so the selected image size should be an integer multiple of 16 at last. In the decoder, we alternated between up-sampling, convolution, and cascade jump connections, and then pass the features learned in the encoding phase directly to the generation registration layers. The successive layer of the decoder on a finer spatial scale can be used to achieve accurate anatomical structure alignment.

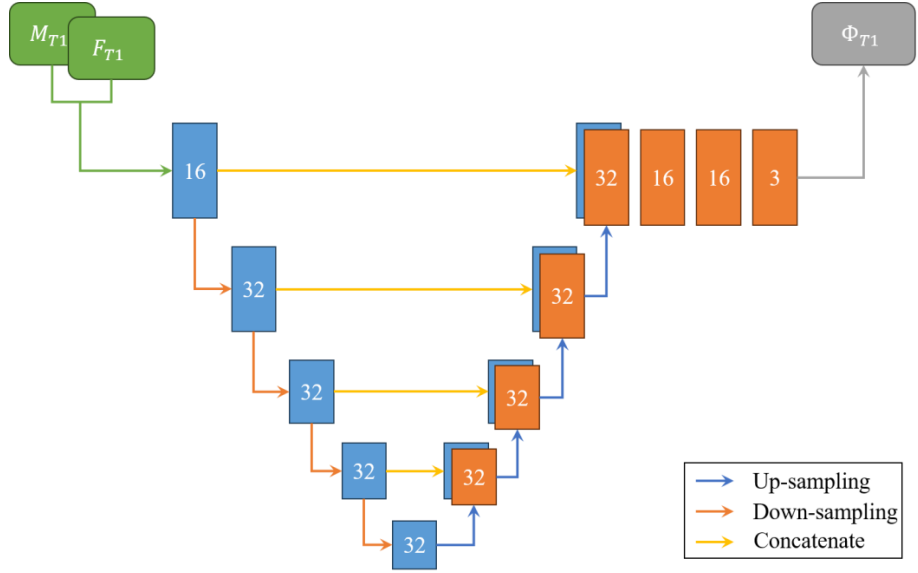

**Fig. 2** The structure of 3D U-Net used in our model. The rectangle in the figure represents the result after convolution operation, and the number in the middle represents the number of channels.

After the input image pair is calculated by the encoder and decoder, the $\phi_{T1}$ (expressed as the displacement field $u_{T1}$) is obtained for the image to be warped in the next step. In addition, the $\phi_{T1}$ is down-sampled to obtain the $\phi_f$ (expressed as the displacement field $u_f$) that can be applied to warp the fMRI in the following steps. Meanwhile, the warped image $R_{T1}$ and $R_f$ are obtained after 3D U-Net and STN. Next, a loss function needs to be constructed to measure the similarity between T1w-MRI and fMRI. Also keeping the deformation field with differential homogeneous nature, a penalty term is needed to maintain the smoothness of the deformation field. After the loss function is available, the optimizer can be used to update the network parameters, and the optimal parameters can be found by minimizing the expected loss function on the training dataset. The following is a concrete implementation of the loss function in our model.

We construct a total loss function $L$ based on T1w-MRI image information and fMRI image information, containing a T1w-MRI similarity loss function, an fMRI similarity loss function, and a regularized loss function for the deformation field $\phi_{T1}$.

$$L = L_{T1-sim} + \lambda L_{f-sim} + \gamma L_{smooth} \tag{3}$$

where $\lambda$ and $\gamma$ are the weight parameters. $L_{T1-sim}$ is used to measure the alignment effect of brain image structure area, $L_{f-sim}$ is used to measure the alignment effect of brain image functional regions, and $L_{smooth}$ is used to regularize the deformation field to make it smoother and more realistic. For each item in the loss function, the specific steps are constructed as follows.

### 2.3.1 Structural loss item

$L_{T1-sim}$ is the similarity loss function between T1w-MRI's warped image $R_{T1}$ and fixed image $F_{T1}$. Generally, it can be calculated using the mean square error (MSE), normalized cross-correlation, the sum of square differences, normalized cross information (NMI) [32], etc. Here, we use the MSE of the grayscale values of images to calculate the similarity of the two images, which is applicable when $F_{T1}$ and $R_{T1}$ have similar image intensity distribution and local contrast, as shown in the following formula:

$$L_{T1-sim} = MSE(F_{T1}, R_{T1}) = \frac{1}{\Omega} \sum_{p \in \Omega} |F_{T1}(p) - R_{T1}(p)|^2 \tag{4}$$

where $p$ is the voxel position and $\Omega$ is the image domain of the whole image

### 2.3.2 Function loss item

The second loss item $L_{f-sim}$ is the similarity function between the warped image $R_f$ and the fixed image $F_f$ of fMRI in the total loss function $L$. A feasible option is extracting valid functional features from the fMRI and aligning them directly or with assistance based on these functional features. Conroy et al. [25] proposed that FC can measure the correlation coefficient of functional signals between two voxels or two brain regions in fMRI data. The closer the FC between two voxels or brain regions at the same location, the more similar their functions. Therefore, the FC features extracted from the whole brain can be used to evaluate the similarity of two fMRI images [25, 33].

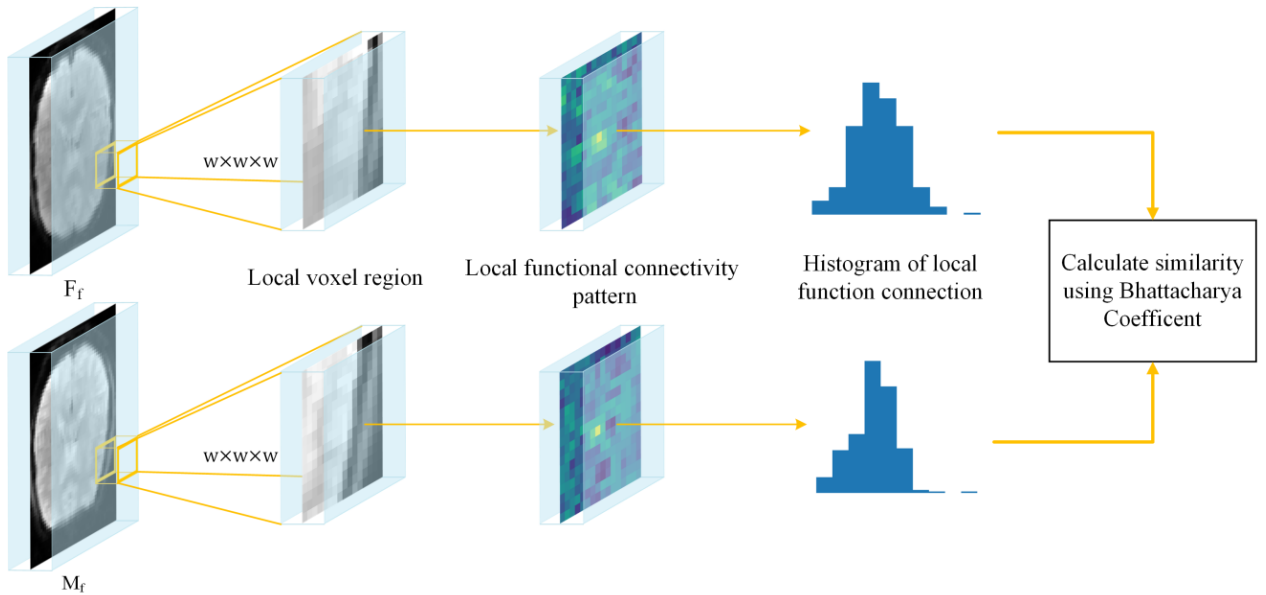

**Fig. 3** A method for calculating fMRI similarity using local functional connectivity patterns.

In addition, the local FC has been calculated in the small space of local neighborhood of each voxel to describe the fMRI function information. Compared with whole-brain FC, using local FC is more robust because whole-brain FC is sensitive to local interference[26], such as small space rotation or shifts of functional units. Therefore, we use the local FC pattern as the feature to calculate fMRI similarity. Specifically, we first calculate the two local area function connectivity, convert it into histogram statistics, and then use the Bhattacharyya coefficient (BC) [34] of the histogram of the two images to calculate the similarity loss, as shown in **Fig. 3**. In this method, the local FC pattern is calculated as follows: For a cubic region in the brain with a side length of $w$ (odd), the Pearson correlation coefficients are calculated between the time series of all voxels within the region and the time series of the central voxel. These correlation coefficients are then utilized to generate three-dimensional correlation maps. For a pair of time series $I(n)$ and $I(m_i)$, their Pearson correlation $C(n, m_i)$ is given :

$$C(n, m_i) = corr(I(n), I(m_i))$$
$$= \frac{\langle I(n) - \overline{I(n)}, I(m_i) - \overline{I(m_i)} \rangle}{\|I(n) - \overline{I(n)}\| \cdot \|I(m_i) - \overline{I(m_i)}\|} \quad (5)$$

Within a cube of the brain with side length $w$ voxels, its central voxel location is $n$ and $m_i$ is the voxel point in the neighborhood of $n$. A 3D FC map is obtained after each voxel is computed with the central voxel.

The resulting correlation plots were next converted to distribution histograms, which were done to eliminate the effect of the relative positions of the voxels. The Pearson correlation coefficient value ranges from -1 to 1, so we divide the range into 21 intervals with an interval of 0.1, and map it to the range of 0 to 20. We count the number of correlation coefficients in each interval in the 3D FC map, and finally get the correlation coefficient distribution histogram, which is the local FC distribution histogram, represented by vectors $\boldsymbol{H}(n) = [H(n, 0), H(n, 1), H(n, 2), \ldots, H(n, 20)]$.

Finally, the difference between the histograms $\boldsymbol{H}_F(n)$ and $\boldsymbol{H}_R(n)$ for the same regions of the two images was calculated using BC. The whole brain is divided into multiple cube regions according to the side length w, and the differences in their histogram are also calculated for all regions $\Omega$. The final loss value is obtained by averaging all the obtained values with the following equation.

$$L_{f-sim} = \frac{1}{\Omega} \sum_{n \in \Omega} BC(\boldsymbol{H}_F(n), \boldsymbol{H}_R(n))$$
$$= \frac{1}{\Omega} \sum_{n \in \Omega} \sqrt{1 - \frac{1}{21\sqrt{\overline{\boldsymbol{H}}_F(n)\overline{\boldsymbol{H}}_R(n)}} \sum_{i=0}^{20} \sqrt{H_F(n, i) H_R(n, i)}} \quad (6)$$

*2.3.3 Spatial smoothing loss item*

In practical situations, physically unrealistic non-smooth deformation fields may be generated, so the deformation fields need to be smoothed to prevent spatial folding. The diffusion regularization is used here to constrain the spatial gradient of the displacement field $u$.

$$L_{smooth}(\phi) = \sum_{p \in \Omega} \|\nabla u(p)\|^2 \quad (7)$$

The spatial gradient can be approximated using the difference of adjacent voxels of the displacement field. Specifically, $\nabla u(p) = (\frac{\partial u(p)}{\partial x}, \frac{\partial u(p)}{\partial y}, \frac{\partial u(p)}{\partial z})$, where the gradient for the x-direction can be approximated by the following equation.

$$\frac{\partial u(p)}{\partial x} \approx u\left((p_x + 1, p_y, p_z)\right) - u\left((p_x, p_y, p_z)\right) \quad (8)$$

The same is true for the y-direction and z-direction.

## 3 Experiment and analysis

*3.1 Dataset*

The dataset was obtained from the OpenfMRI database. Its accession number is ds000030 [35]. We had a total of 244 subjects after manually removing the unqualified images, which contained 244 T1w-MRI data and the corresponding 244 resting-state fMRI data. Participants ranged in age from 21 to 50, with an average age of 33, including 144 males and 100 females. The dataset includes healthy individuals (112), schizophrenic (48), bipolar disorder (47), and attention deficit hyperactivity disorder (37). All data is collected by Siemens Trio. Each subject's functional image consisting of 152 time points with a repetition time (TR) of 2000.0 ms; 34 slices were imaged. The pixel spacing in the X and Y dimensions was 3 mm and the slice thickness was 4 mm.

*3.2 Preprocessing*

Firstly, all T1w-MRI data in the dataset were preprocessed using Freesurfer [36]. Specific steps include skull removal, intensity standardization, and subcortical structure segmentation, in which the subcortical structure segmentation labeled image (Seg) was used for the subsequent evaluation of the results, and then the T1w-MRI and Seg were simultaneously affine aligned to a standard spatial template of MNI152 with a voxel size of 1 mm using SPM software[5]. Subsequently, the T1w-MRI and Seg were cropped to a size of 144 × 192 × 192, and the T1w-MRI images were normalized to a gray scale between 0 and 1. The fMRI data were preprocessed with DPARSF [6] as follows: (1) Removing the first 10 time points. (2) Slice timing. (3) Head motion correction (4) Nuisance covariates regression (use the Friston 24). (5) Detrend. (6)

Band-pass filtering of 0.01–0.1 Hz. (7) Spatial smoothing (FWMH:[4,4,4]) were applied to the functional image. The fMRI images were affinity aligned to the MNI152 template using SPM. Specifically, the mutual information between the fMRI average image and MNI152_3mm is used to compute the affine matrix. Subsequently, this affine matrix is applied to the fMRI images to affine-align them to the MNI space, ensuring they match the same dimensions, voxel sizes, orientation, etc., as the MNI152_3mm template. The fMRI images were cropped to a size of $48 \times 64 \times 64 \times 142$. Time point was 142 after preprocessing the dataset used in this experiment. Finally, the processed data pairs were split into an 8:1:1 ratio, 194 as the training set, 25 as the validation set, and 25 as the test set.

*3.3 Implementation*

The registration experiments are generally divided into inter-subject registration and atlas-based registration. Atlas-based registration is a common formulation in population analysis, where inter-subject registration is a core problem [17]. In all experimental stages, two participants are randomly selected from the training set for training in each step.

The registration between different subjects has more variability, and the model faces more complex situations. In this experiment, one subject in the test set is used as a reference ($F_{T1}$ and $F_f$), and the other subjects are sources ($M_{T1}$ and $M_f$).

Atlas-base registration holds practical significance. In fMRI group analysis, individual subjects are typically registered to a standard template. In this experiment, all test data ($M_{T1}$ and $M_f$) are registered to the MNI152 space ($F_{T1}$ and $F_f$), and the results are analyzed accordingly.

In order to obtain the optimal model performance, we first discuss the effect of cubes at different functional area scales, i.e., with different side lengths w, and restrict the weight parameter of deformation field regularization such that γ= 0.01 to control the variables. The experimentally obtained optimal value of $w$ is then used to discuss the effect of the functional similarity loss weight $\lambda$. The above two steps are mainly experiments on the influence of hyperparameters, followed by experiments to evaluate our final model and several baseline methods on the test set. The main baseline methods for comparison are VoxelMorph, Transmorph, and Syn:

Affine: All data are registered to MNI space through an affine transformation during preprocessing, so we can use the processed data as a control group to compare our lifting effect.

Voxelmorph [17]: Our model is an improvement of the Voxelmorph model and therefore serves as the most important comparison method. Here we use the unsupervised model of Voxelmorph as the baseline method, using the loss function MSE and the optimal parameter value $\lambda = 0.02$.

Transmorph [18]: This method can be seen as an improvement of the Voxelmorph, where the Swin-Transformer is used as the encoder to replace the encoder in the U-Net. We use MSE as the loss function and set $\lambda = 0.02$。

Syn [13]: Syn is a nonlinear registration algorithm in ANTs [7]. We use the mean squared difference as the objective function, and the default Gaussian smoothing of 3 to perform three scales of 180, 80, and 40 iterations respectively in the brain MRI registration task.

Each of these models can obtain the deformation parameters by registering the T1w-MRI image pairs, and then imposing them on the fMRI images. Experiments are conducted in the same training set, verification set, and test set. In the training stage, for the model based on deep learning, images of a pair of subjects are randomly selected from the training set for training. Traditional algorithms do not require training. In the testing and validation stage, to facilitate group analysis of fMRI, we register the other subjects in the group to one of the subjects.

Our model uses Tensorflow [37] as the backend Keras. The model is trained using the Adam optimizer with a learning rate set to $10^{-4}$ and 500 iterations, with 100 steps in each generation. The deep learning-based model is computed using graphics card acceleration with an NVIDIA 2080 device with 12G of video memory. Other traditional optimization methods use CPU computation.

*3.4 Valuation metrics*

*3.4.1 Dice score*

Dice score [38] is used to assess the overlap of brain anatomical structures. A good registration algorithm should make the brain anatomical structures of the two subjects align as much as possible. When T1w-MRI was preprocessed using Freesurfer, anatomical region segmentation was also performed by taking the regions containing at least 100 voxels, and a total of 30 anatomical regions [17] were selected. The warped Seg was obtained by applying the same deformation field $\phi_{T1}$ to the segmented images of T1w-MRI. The Dice score of the k region is calculated as the ratio of the overlapping volume to the total volume in the k anatomical region of the warped Seg $s_r^k$ and the same anatomical region of the fixed Seg $s_f^k$. The final Dice evaluation index can be obtained by averaging the Dice scores of all regions separately.

$$Dice = \frac{1}{30}\sum_{k=1}^{30}\left(2\cdot\frac{|s_f^k \cap s_r^k|}{|s_f^k| + |s_r^k|}\right) \tag{9}$$

*3.4.2 Group-level t-map*

Since there is no gold standard for directly assessing the performance of functional registration, the evaluation of fMRI registration performance is usually based on the group-level statistical map of resting-state brain functional networks [39], in which independent component analysis (ICA) [40] can be used to extract resting-state brain networks. For the purpose of evaluation, four brain networks included default mode network (DMN), visual network (VN), central executive network (CEN), and sensorimotor network (SMN) are selected and extracted by using group ICA [41] in fMRI Toolbox (GIFT) [42]. In particular, ICA was performed 100 times using ICASSO at different initial settings to generate 20 independent components on the test set, and then selected the above four common group network components from them.

Next, Z-score transformation is applied to the values in the group network components, and a one-sample t-test was used to examine the degree of co-activation of each voxel point within the group, resulting in a group-level t-test map (t-map). A higher degree of co-activation indicates better functional correspondence between individuals. During the adjustment of hyperparameters, the average of the peak values (peak-value) on the t-map corresponding to the four network components is used to indicate the fMRI similarity of a group of subjects. In addition, in the comparative experiments, we analyzed the results by setting a threshold and using the number of voxels exceeding the threshold for analysis.

## 4 Results

### 4.1 Hyperparameter selection

In this section, we observe the effect on the alignment results by adjusting the hyperparameters of the model and obtaining a better hyperparameter value.

**Fig.4** shows the influence of FC patterns on the mean Dice score and peak-value on the validation set for different scales w. When the weight parameter in the loss function is set to $\lambda = 0.01$ and $\gamma = 0.01$, we tested the results of the model on the validation set with $w$ is 3, 5, 7, 9, 11, 13, 15, 17, 21, 23, 41 and 47. The results show that the peak-value increases gradually as the scale $w$ of the functional regions increases. The Dice scores first increased with increasing $w$ and then gradually decreased, reaching the maximum at $w$=21. Therefore, $w$ =21 is chosen to train our deep learning network and evaluate our method. In **Fig.4** (a), the alignment of FC patterns using smaller scales or global scales is not good enough to maintain the alignment of structural regions and functional regions at the same time. Therefore, the FC pattern contributes to the alignment effect at a certain scale, but beyond a certain range, it leads to the destruction of the structural regions.

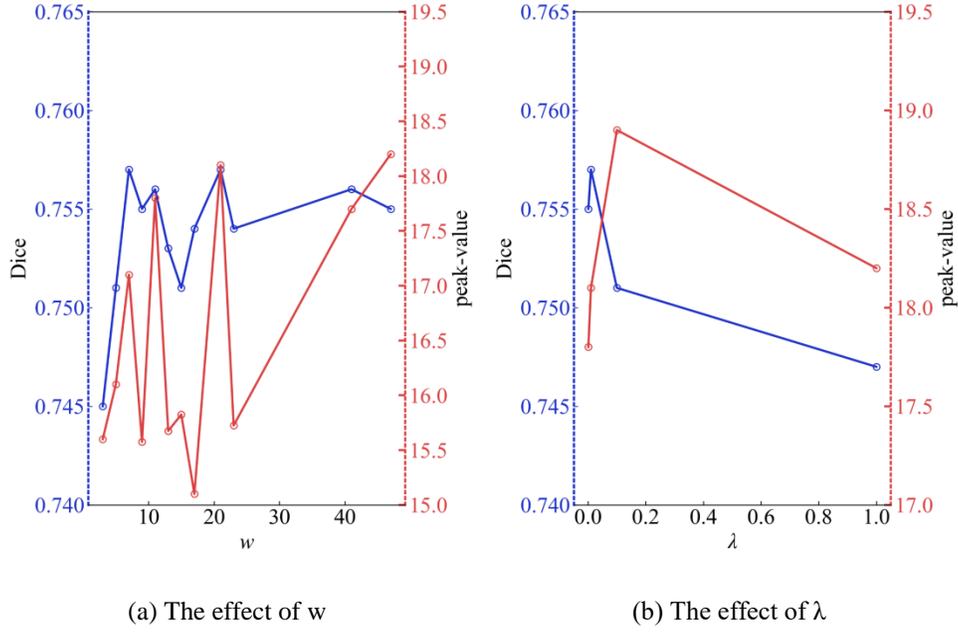

(a) The effect of w  (b) The effect of λ

**Fig. 4** The effect of hyperparameters w and λ on Dice and peak-value of registration results.

In addition, **Fig.4** also shows the effects of different function loss weight parameters $\lambda$ on the registration results at 0.001, 0.01, 0.1, and 1. As the weight of parameter $\lambda$ gradually increases, the average peak-value of four function networks increases and peaks at $\lambda = 0.1$, and then gradually decreases. The Dice scores also gradually increase and then decrease, reaching a peak at $\lambda = 0.01$. In order to keep the structural region stable and improve the functional region, $\lambda = 0.01$ is chosen to train our deep learning network and evaluate our method. In **Fig.4** (b), the size of the functional loss weight coefficient $\lambda$ determines the proportion of the functional image alignment effect to the total effect. As the weight $\lambda$ increases, the proportion of functional similarity loss gradually becomes larger, and its alignment effect is improved. Since fMRI and T1w-MRI come from the same brain, their information in the same region can complement each other, so it will promote the overall registration effect to a certain extent. However, the model will focus too much on the registration effect of the fMRI after the weight $\lambda$ exceed a certain value, leading to a decrease in the T1w-MRI, which in turn leads to a decrease in the overall registration effect.

*4.2 Inter-subject registration*

In this experiment, we test our model for inter-subject registration. We select $w = 21$ and $\lambda = 0.01$ to train the model and conduct comparative experiments with other baselines.

**Fig. 5** shows the registration results of sample T1w-MRI in the test set for our model and several baseline methods, including VoxelMorph, Transmorph, and Ants-Syn. The $R_{T1}$ in the figure is all close to $F_{T1}$ overall with some differences in the detail parts. The deformation fields generated by the model are also visualized in the figure using RGB and grid forms, and it can be seen that Ant-Syn shifts more voxel points to a greater extent. After the corresponding deformation fields were obtained by registering the T1w-MRI, we compared the corresponding fMRI registration results of the baseline methods in the test set.

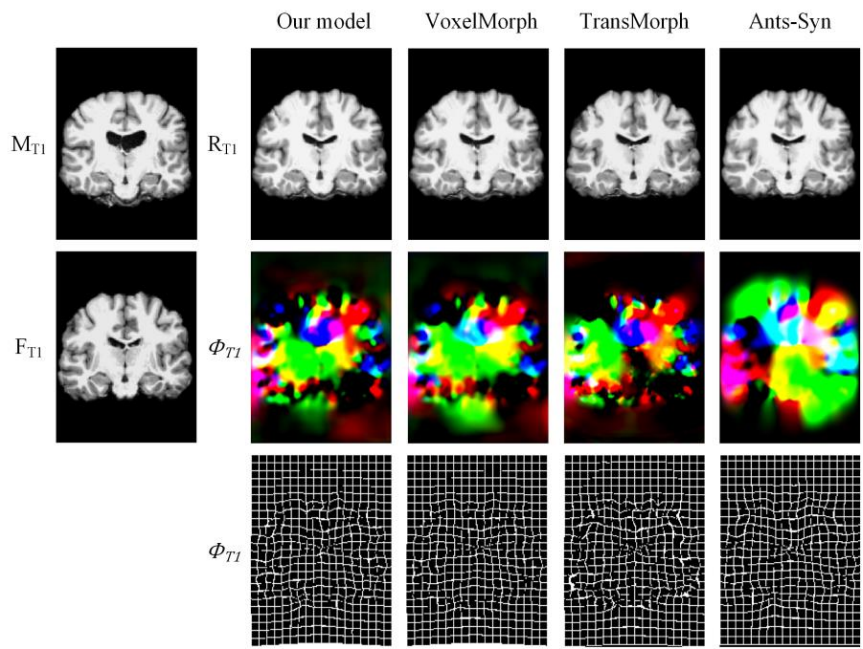

**Fig. 5** Example results of Our model, VoxelMorph, TranMorph and Ants-Syn methods for registering a pair of T1w-MRI ($M_{T1}$ and $F_{T1}$) to obtain a warped image ($R_{T1}$) and deformation fields ($\phi_{T1}$). Displacement in each spatial dimension is mapped to each of the RGB color channels in column 4.

**Fig. 6** shows an example of fMRI registration results of these methods in the test set, $M_f$, $F_f$, and $R_f$ images are displayed as the coronal view at the same time point. From the example, it can be seen that Ants-Syn produces a deformation field with a large shift that over-distorts the fMRI.

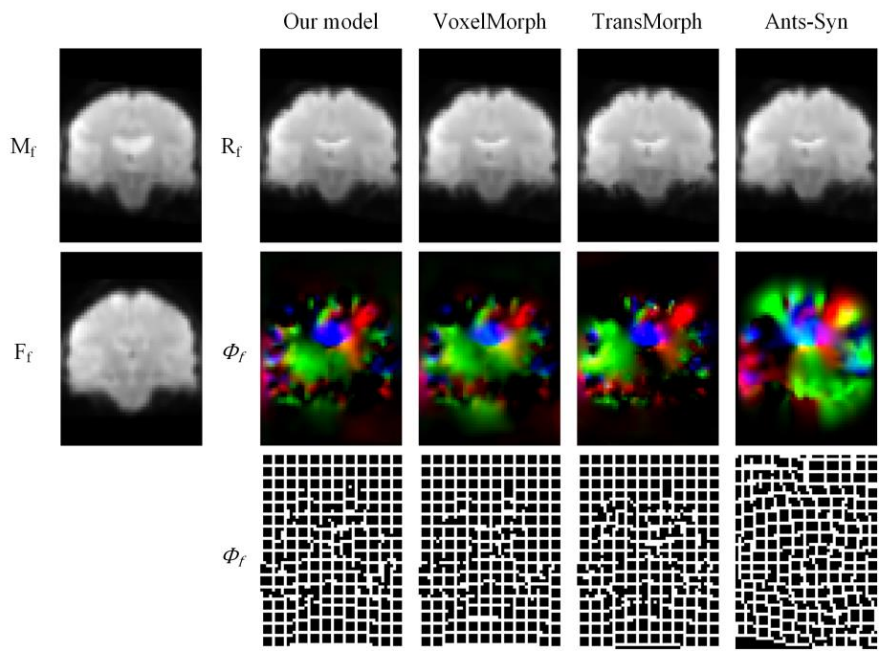

**Fig. 6** Example results of Our model, VoxelMorph, TranMorph and Ants-Syn methods for registering a pair of fMRI ($M_f$ and $F_f$) to obtain a warped image ($R_f$) and deformation fields ($\phi_f$).

By selecting different thresholds T, we computed the number of voxels greater than T in the t-maps of the DMN, CEN, SMN, and VN networks. We then plotted the average voxel count relative to the T, as shown in **Fig. 7**. It can be seen that when T is small, our model has a high average area of co-activated regions, indicating that our model has a larger functional region alignment. Notably, as the threshold increases, our model is still at the highest level with a high degree of co-activation, although the area of the region decreases.

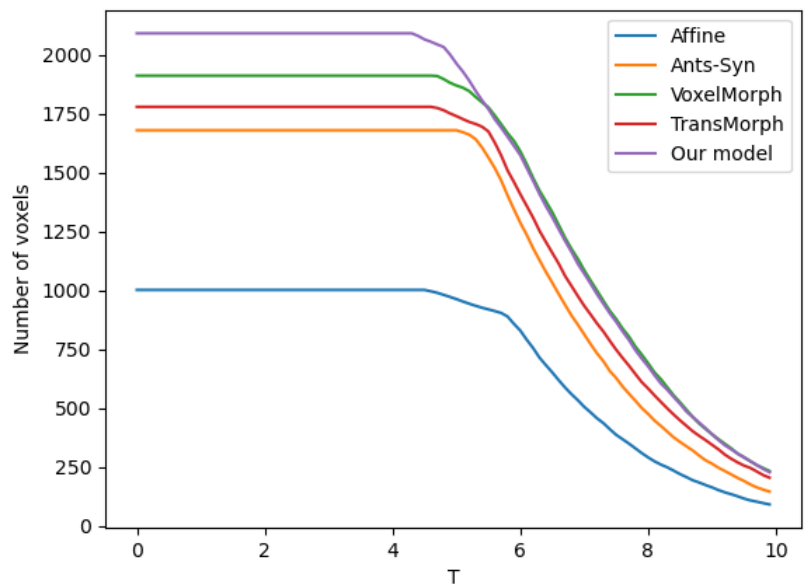

**Fig. 7** The average number of voxel over thresholds T in DMN, CEN, SMN and VN networks.

**Table I** illustrates the count with t-map values exceeding 4.24 ($p < 0.05$) for the four network components following the registration of images in the test set. In the result, the affine transformation has the smallest number. Our model has the largest number in DMN, CEN and VN networks. On average, the number of voxels exceeding the threshold in our model reached 2248.00, which was the largest among all comparison methods.

TABLE I Comparison of the speed of inter-subject registration and the voxel number of the co-activation area at T=4.24 using different methods (P<0.05)

| Method | DMN | CEN | SMN | VN | Average | Time(s) |
|---|---|---|---|---|---|---|
| Affine | 1058 | 1039 | 545 | 1368 | 1002.50 | - |
| Ants-Syn | 276 | 1153 | **2667** | 1454 | 1387.50 | 252.61 |
| Voxelmorph | 1973 | 1571 | 2616 | 2076 | 2059.00 | 2.80 |
| Transmorph | 1763 | 1438 | 2390 | 1941 | 1883.00 | 2.86 |
| Our model | **2044** | **2155** | 2590 | **2203** | **2248.00** | **2.16** |

Abbreviations: DMN: Default Mode Network; CEN: Central Executive Network; SMN: Sensorimotor Network; VN: Visual Network.

Subsequently, the regions obtained in the previous step were compared with the DMN, SMN, CEN, and VN brain networks from the Smith BrainMap atlas [39] to assess region overlap. Among these four components, the size of overlap between regions exceeding the threshold of T=3 in the Smith atlas and the regions obtained in our previous step was calculated. The results are presented in **Table II**. A decrease in voxel count indicates that some regions are not present in the Smith atlas. The average voxel count of intersecting regions obtained by our model is 1405.00, indicating that these brain networks are closest to the brain network regions in the Smith atlas.

TABLE II Comparison of the overlap voxel count between the co-activated regions obtained from inter-subject registration and the brain networks in the Smith Atlas.

| Method | DMN | CEN | SMN | VN | Average |
| --- | --- | --- | --- | --- | --- |
| Affine | 742 | 719 | 468 | 985 | 728.50 |
| Ants-Syn | 214 | 904 | **1769** | 987 | 1060.50 |
| Voxelmorph | 954 | 1083 | 1585 | 1644 | 1311.00 |
| Transmorph | 916 | 1051 | 1421 | 1520 | 1227.00 |
| Our model | **959** | **1312** | 1592 | **1757** | **1405.00** |

**Fig. 8** shows the group-level t-maps of four networks with T=4.24 when all methods are used based on inter-subject registration.

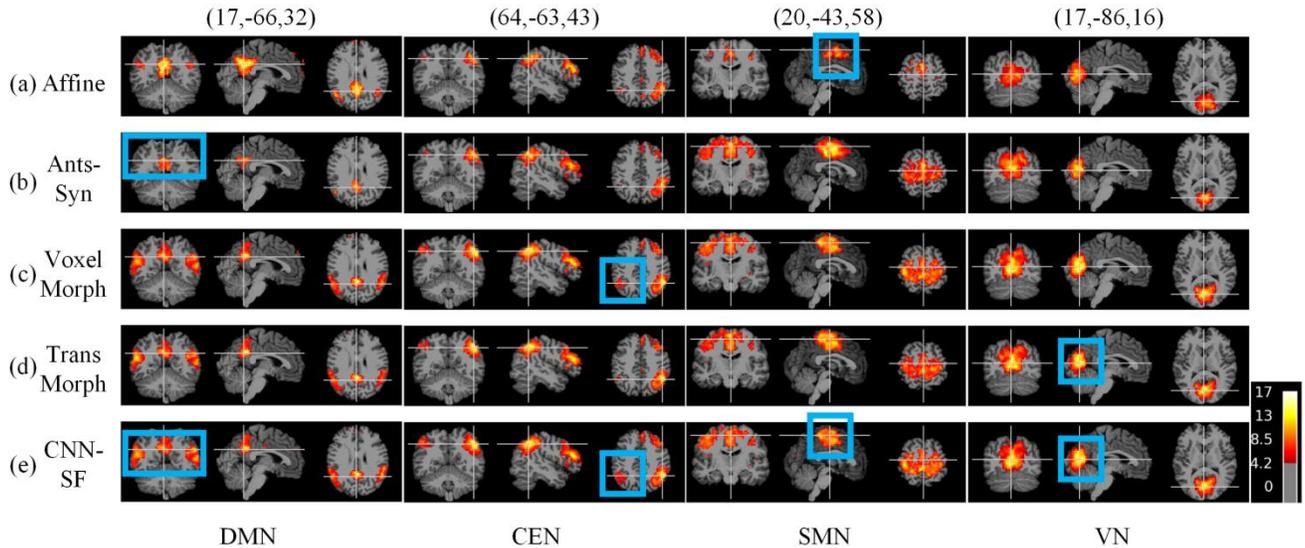

**Fig. 8** Group-level t maps of DMN, CEN, SMN, and VEN over the threshold T=4.24 (p < 0.05) after inter-subject registration by (a) Affine (b) Our model (c) VoxelMorph (d) TransMorph (e) Ants-Syn. The p-value was corrected using the False Discovery Rate (FDR).

*4.3 Atlas-base registration*

In this experiment, we test our model for Atlas-base registration. Specifically, the T1w-MRI is first registered to the MNI152 template to obtain a deforming field. Subsequently, the functional images of individual subjects are linearly registered to their respective T1w images. Finally, the obtained deformation field is utilized to register the fMRI data to the MNI152 space.

**Table III** displays the number of voxels with t-map values exceeding 4.24 ($p < 0.05$) for each of the four network components after registration of the test set subjects to the standard template. The results indicate that the fewest number of voxels underwent affine transformation. Our model exhibits the highest count in the DMN, CEN, and VN networks. The average voxel count exceeding the threshold reached 3620, which remains the highest among all comparative methods.

**TABLE III** Comparison of the voxel number of the co-activation area at T=4.24 using different methods based on atlas registration (P<0.05).

| Method | DMN | CEN | SMN | VN | Average |
|---|---|---|---|---|---|
| Affine | 1058 | 1039 | 545 | 1368 | 1002.50 |
| Ants-Syn | 2478 | 2427 | **5495** | 2800 | 3300.00 |
| Voxelmorph | 1288 | 1895 | 3773 | 2628 | 2396.00 |
| Transmorph | 1071 | 1800 | 4446 | 2858 | 2543.75 |
| Our model | **2489** | **3448** | 5476 | **3067** | **3620.00** |

**Table IV** shows the size of the overlap between the regions exceeding the threshold of T=3 in the DMN, SMN, CEN, and VN brain networks of the Smith atlas and the regions obtained in our previous step. The voxel count for the obtained regions has decreased in all cases. The average voxel count of intersecting regions obtained by our model is 1846.00, which remains the closest to the brain network regions in the Smith atlas.

**TABLE IV** Comparison of the overlap voxel count between the co-activated regions obtained from atlas-base registration and the brain networks in the Smith Atlas.

| Method | DMN | CEN | SMN | VN | Average |
|---|---|---|---|---|---|
| Affine | 742 | 719 | 468 | 985 | 728.50 |
| Ants-Syn | **1090** | 1533 | 2401 | 1570 | 1648.50 |
| Voxelmorph | 665 | 1165 | 1776 | 1816 | 1355.50 |
| Transmorph | 553 | 1188 | 2114 | 1851 | 1426.50 |
| Our model | 1028 | **1951** | **2451** | **1954** | **1846.00** |

**Fig. 9** shows the group-level t-maps of four networks with T=4.24 when all methods are used based on atlas-base

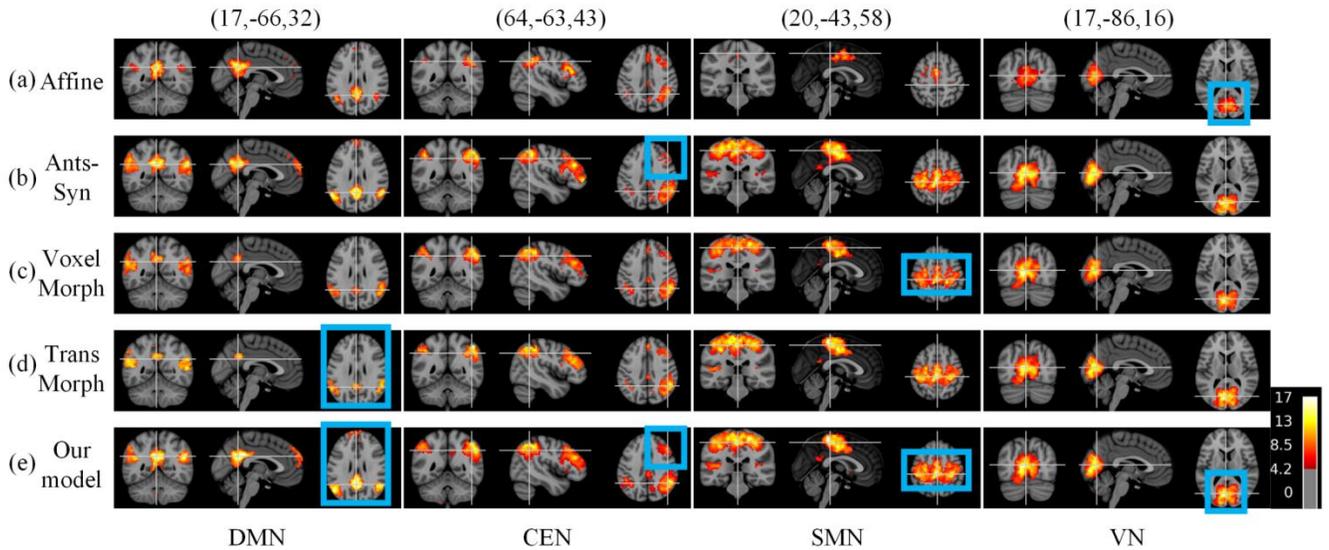

**Fig. 9** Group-level t maps of DMN, CEN, SMN, and VEN over the threshold T=4.24 (P<0.05) after atlas-base registration by (a) Affine (b) Our model (c) VoxelMorph (d) TransMorph (e) Ants-Syn. The p-value was corrected using the False Discovery Rate (FDR).

## 5 Discussions

The registration model proposed in this paper demonstrates superior performance compared to baseline methods, effectively achieving better alignment of brain functional regions and ensuring reliable fMRI registration. Despite yielding satisfactory results in fMRI registration, our method still has some limitations. In VN, CEN, and DMN, our approach shows larger functional co-activation areas compared to other methods. While the proposed model is effective in aligning certain brain functional regions, such as those in VN, CEN, and DMN, it may not exhibit the same advantages in individual brain areas like the SMN. Furthermore, while the voxel count in co-activated regions of the four brain networks used as an evaluation metric for fMRI in this study yielded promising results, it also has significant limitations. It cannot provide a comprehensive assessment of the alignment level of functional regions from a holistic perspective. Therefore, further exploration of gold standards more suitable for fMRI alignment is paramount.

In exploring the scale size of local functional connectivity patterns and the weight parameters of functional loss functions, this study has identified values that are relatively suitable. Subsequent work may consider conducting a global-scale search or employing more effective parameter search methods to obtain optimal values.

Additionally, during registration, the model can achieve favorable results by leveraging T1w-MRI. In future research, we will consider using fMRI as the primary modality and structural MRI as auxiliary information to generate deformation fields. Such an improvement can address situations where structural images are lacking.

## 6 Conclusions

Functional MRI registration serves as the foundation for group-level analysis, holding significant importance in functional localization and group difference analysis. In this study, we propose a CNN-based registration model that integrates structural and functional MRI information. Experimental results demonstrate that our model outperforms other algorithms in both inter-subject and atlas-based registrations, yielding superior registration accuracy. Specifically, our model exhibits improvements over the baseline model Voxelmorph. Thus, the incorporation of functional information can enhance functional alignment to some extent. From the perspective of fMRI registration, it solves the drawback of existing fMRI registration algorithms based on deep learning not fully utilizing structural information. Moreover, we also propose an fMRI local functional connectivity pattern to promote functional consistency. It uses statistical information, which can eliminate the influence of location factors. In addition, the framework based on deep learning can improve the computing speed compared with the traditional algorithm. This model has the potential to optimize fMRI image processing. It is helpful to promote the development of fMRI research and application.

## Acknowledge

This work was supported in part by the National Natural Science Foundation of China under Grants 61906117 and 62276164, and in part by the `Science and technology innovation action plan' Natural Science Foundation of Shanghai under Grant 22ZR1427000.